\documentclass[a4paper]{jpconf}

\usepackage{citesort}

\usepackage{graphicx}

\usepackage{amsmath,amssymb}

\newcommand{\beq}{\begin{equation}}
\newcommand{\eeq}{\end{equation}}

\begin{document}
\title{Towards Quantum Transport
for Central Nuclear Reactions}

\author{Pawel Danielewicz}

\address{National Superconducting Cyclotron Laboratory, Michigan State University, East Lansing, MI 48824, USA}

\ead{danielewicz@nscl.msu.edu}

\author{Arnau Rios}

\address{Department of Physics, University of Surrey, Guildford, GU2 7XH, United Kingdom}

\ead{a.rios@surrey.ac.uk}

\author{Brent Barker}

\address{Biological, Chemical and Physical Sciences Department, Roosevelt University, Chicago, IL 60605, USA}

\ead{bbarker@roosevelt.edu}

\begin{abstract}
Nonequilibrium Green's functions represent a promising tool
for describing central nuclear reactions.  Even at the single-particle level, though, the Green's functions contain more information that computers may handle in the foreseeable future.  In this study, we explore slab collisions in one dimension, first in the mean field approximation and demonstrate that only function elements close to the diagonal in arguments are relevant, in practice, for the reaction calculations.  This bodes well for the application of the Green's functions to the reactions.  Moreover we demonstrate that an initial state for a reaction calculation may be generated through adiabatic transformation of interactions.  Finally, we report on our progress in incorporating correlations into the dynamic calculations.

\end{abstract}

\section{Introduction}

Expansion in the scope of nuclear investigations is largely owed to the construction of new accelerator facilities, in particular those accelerating secondary exotic beams, leading to the growth in discovered nuclides and nuclides that can be used in further experimentations.  A~second high-intensity generation of the latter facilities is now under construction and these include the Facility for Rare Isotope Beams (FRIB) at the Michigan State University, at the moment the largest investment in the low-energy nuclear physics in North America.  The expansion on the experimental side is paralleled by a growth in the theory for nuclear structure that, till now largely phenomenological, has increasingly become fundamentally based, even down to QCD, as an effective theory.  Not surprisingly, expectations rise then with respect to the reaction theory.  The theory for central reactions has been till now phenomenological to a much greater extent than the structure theory and, in particular, largely semiclassical.  Many particles participate in central reactions, so the theory needs to be statistical in nature.  For dynamics, the natural general way forward is the quantum transport.

\section{Description of Central Nuclear Reactions}

\begin{figure}[ht]
\centerline{
\includegraphics[width=.62\linewidth]{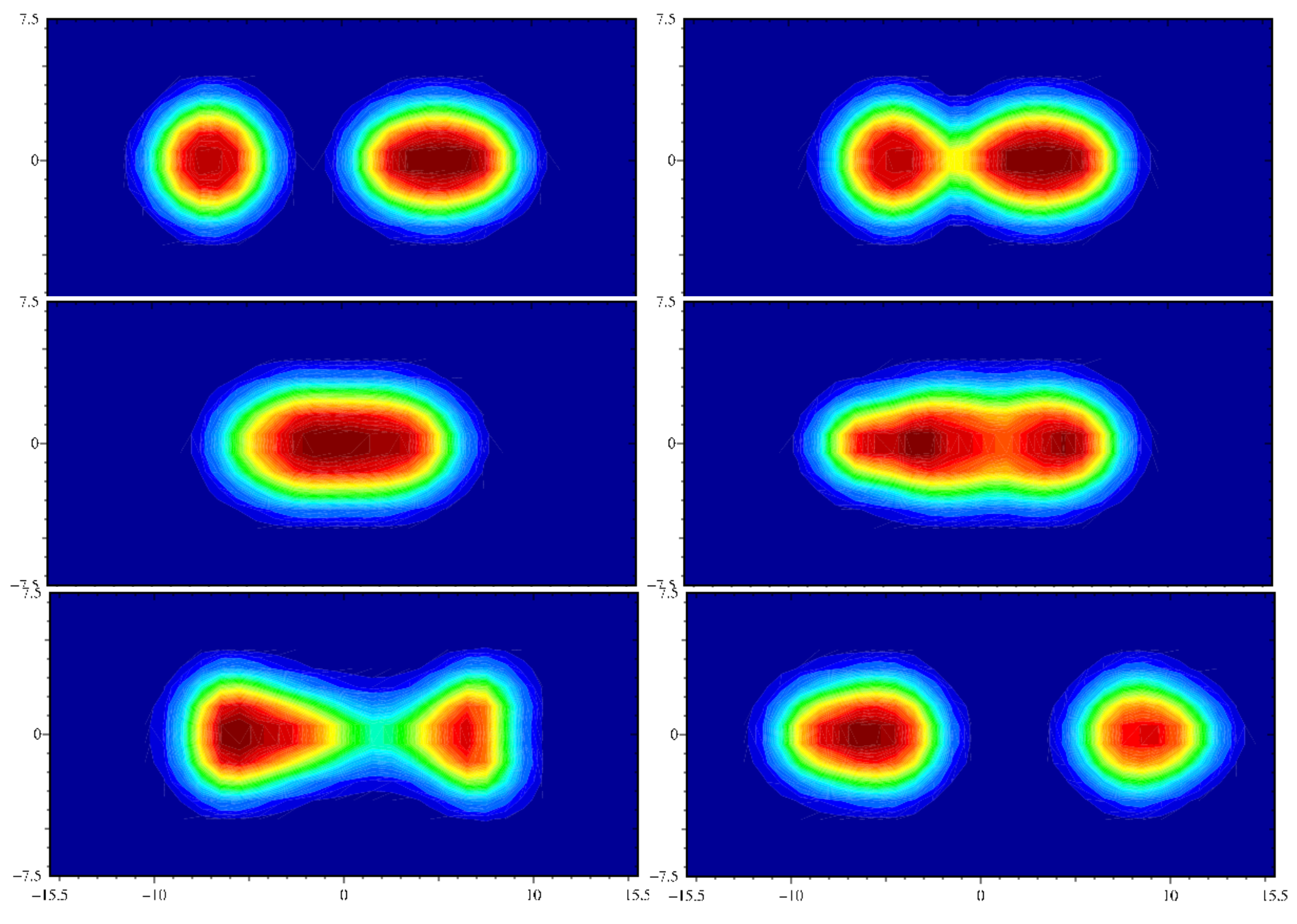}
}
\caption{
Head-on collision of $^{16}$O + $^{22}$Ne at $E_\text{cm} = 95 \, \text{MeV}$ in the TDHF calculation by Umar and Oberacker~\cite{Umar06}.  In head-on collisions at high enough energies, the nuclei pass through each other within TDHF.
}
\label{fig:TDHF}
\end{figure}

Historically, just a handful of methods have been developed to describe central nuclear
reactions.  The time-dependent Hartree-Fock (TDHF) method has been exhaustively employed in describing
low-energy reactions \cite{Negele82}. Nowadays, the TDHF simulations can be performed in full 3D and involve
nuclei as heavy as Uranium \cite{Golabek09}. However, the validity of TDHF requires correlations to play a
negligible role in the dynamics \cite{Tohyama87}. At low energies, one might argue that the role of correlations
is minimized by wavefunction antisymmetrization. Conversely, one would expect correlations to dominate at
higher energies, where role of the Pauli principle weakens.  With increase of collision energy the limitation of TDHF is exhibited in the fact that nuclei in TDHF simulations become transparent to each other, see Fig.~\ref{fig:TDHF}.  No such transparency is observed experimentally.

\begin{figure}
  \centerline{\includegraphics[width=.55\textwidth]{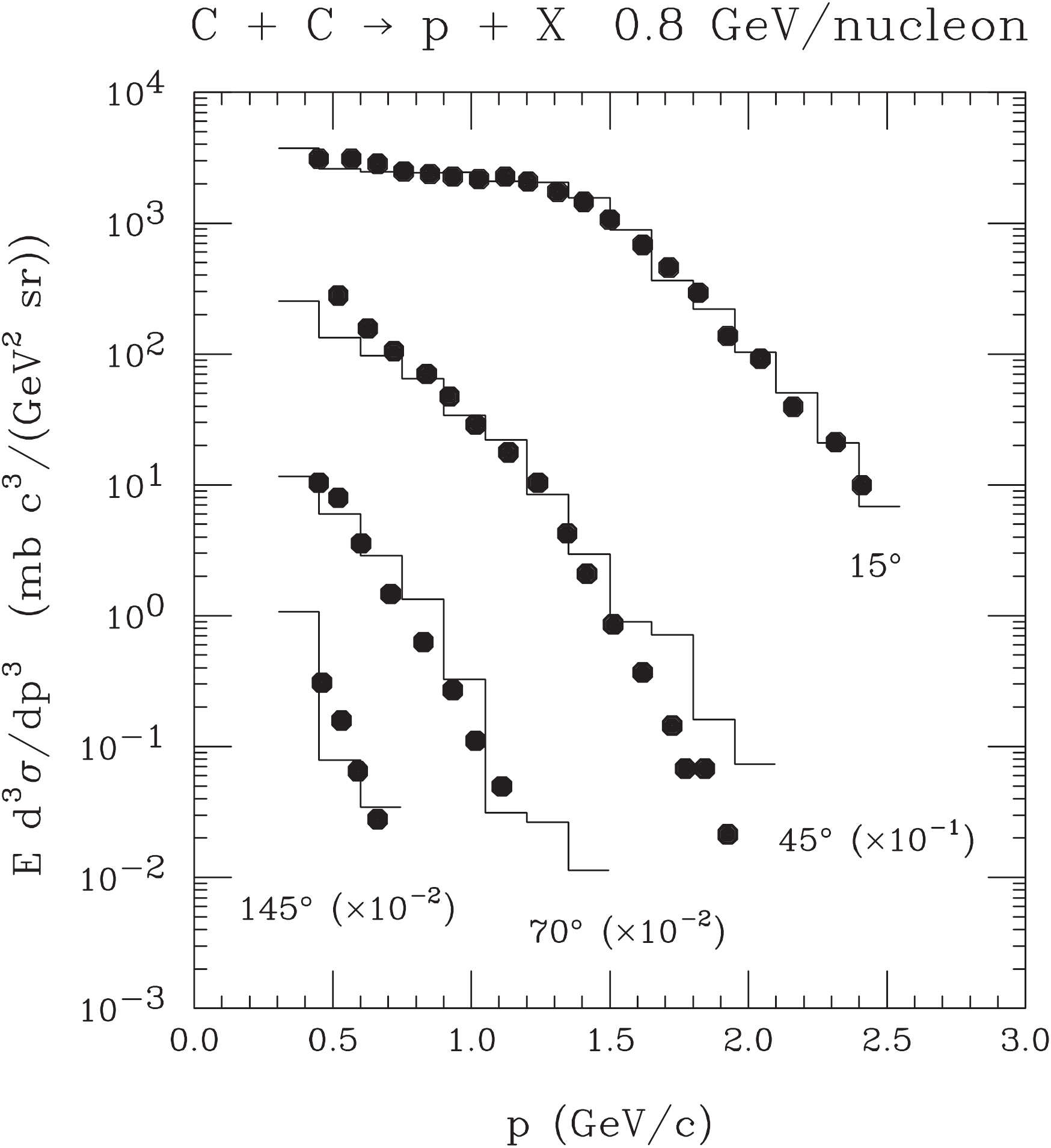}}
  \caption{Proton spectra from an 800 MeV/nucleon $^{12}$C + $^{12}$C reaction.  Dots represent data of Ref.~\cite{nagamiya81} and histograms represent BE calculations of Ref.~\cite{dan91}.}
  \label{fig:CCp}
\end{figure}

Intermediate and high energy central nuclear reactions have been commonly described in terms of the Boltzmann-equation (BE) models \cite{bertsch88}.  In these models the evolution of the phase-space distribution functions $f({\pmb r},{\pmb p},t)$ of nucleons and other particles is followed.  BE models have been fairly successful in describing many aspects of higher-energy reactions, see e.g.~Fig.~\ref{fig:CCp}.  However, the use of BE in reactions has been criticized on theoretical grounds.  BE relies on the quasiparticle picture and simple estimates \cite{danielewicz84} indicate that particle scattering rates are comparable to particle energies, which undermines that picture.  In this context, it is theoretically difficult~\cite{dickhoff98} to separate collisional effects, described with cross-sections and entering the collisional integrals in BE from mean-field effects entering the quasiparticle energies.

\section{Kadanoff-Baym Equations}

The TDHF and BE approaches to central nuclear reactions are fundamentally tied with the single-particle density matrix
\beq
\rho({\pmb r}_1 \, {\pmb r}_{1}' \, t)
= \langle \Phi | \psi^\dagger ( {\pmb r}_1' \, t) \, \psi ( {\pmb r}_1 \, t) | \Phi \rangle \, .
\label{eq:GF<}
\eeq
The density matrix yields all single-particle observables.  E.g.\ the particle density represents the diagonal of this matrix as the expectation value of the density operator:
\beq
n({\pmb r} \, t) = \rho ({\pmb r} \, {\pmb r} \, t) = \langle \Phi | \psi^\dagger ( {\pmb r} \, t) \, \psi ( {\pmb r} \, t) | \Phi \rangle \, .
\eeq
The Wigner function, that may be viewed as the quantal version of the classical phase-space distribution, results from evaluating a Fourier transformation of the equal-time Green's function, i.e.\ density matrix, in relative coordinates:
\beq
f({\pmb p} \, {\pmb r} \, t) = \int \text{d} ({\pmb r}_1 - {\pmb r}_1') \, \text{e}^{-i {\pmb p} ({\pmb r}_1 - {\pmb r}_1')} \, \rho({\pmb r}_1 \, {\pmb r}_{1}' \, t) \, .
\eeq
Here, the spatial argument of the Wigner function is ${\pmb r} = ({\pmb r}_1 + {\pmb r}_{1}')/2$.  It can be seen that the doubled spatial argument in the density matrix accounts for position and momentum in the classical limit.  By extension, one can expect that the doubled time argument in the Green's function corresponds to time and energy in the classical limit, i.e.\ the density in momentum {\em and} energy at a given position and time should be:
\beq
-iG^< ({\pmb p} \, \epsilon \, {\pmb r} \, t) =
\int \text{d} ({\pmb r}_1 - {\pmb r}_1') \, \text{d} (t_1 - t_1') \,
\text{e}^{i[\epsilon (t_1 - t_1')- {\pmb p} ({\pmb r}_1 - {\pmb r}_1')]} \,
(-i) G^<({\pmb r}_1 \, t_1 \, {\pmb r}_{1}' \, t_1') \, .
\label{eq:SpectralF}
\eeq
Indeed, for the static case of a Hartree-Fock state, where the Green's function is generally a superposition of products of occupied orbitals $\phi_\alpha$,
\beq
-iG^<({\pmb r}_1 \, t_1 \, {\pmb r}_{1}' \, t_1') = \sum_\alpha \phi_\alpha ({\pmb r}_1 \, t_1) \,
\phi_\alpha^* ({\pmb r}_1' \, t_1') \, ;
\eeq
Eq.~\eqref{eq:SpectralF} yields
\beq
-iG^< ({\pmb p} \, \epsilon \, {\pmb r} \, t) =  \sum_\alpha f_\alpha({\pmb p} \, {\pmb r}) \, \delta(\epsilon - \epsilon_\alpha) \, .
\eeq
Here, $\epsilon_\alpha$ are single-particle energies of the orbitals.  In nuclear physics, the spectral function represented by Eq.~\eqref{eq:SpectralF} is explored for ground-state nuclei in inelastic electron scattering.

Outside of the mean-field approximation, an intrinsically consistent dynamics for the Green's function \eqref{eq:GF<}, in nonequilibrium or even finite temperature situation, can only be arrived when considering different orderings of single-particle operators in an expectation value, encompassed in the Green's function
\beq
i \, G(1,1') = \langle \Phi | T \left\lbrace \psi (1) \, \psi^\dagger (1')   \right\rbrace    | \Phi \rangle \, .
\label{eq:GF}
\eeq
Here, the time arguments of the operators are assigned to either side of a time contour that runs first forward and then backward in time, representing the evolution of the ket in the expectation value, on one hand, and of the bra, on the other \cite{danielewicz84,kadanoff}.  The superoperator $T$ orders operators according to their order on the time contour.  Depending on the assignment of time arguments to the contour branches, different actual ordering of the operators can result, with the superscript '$<$' in the Green's function referring to the order in \eqref{eq:GF<} and '$>$' to the opposite order.

A perturbative expansion of the evolution operators in the Green's function \eqref{eq:GF}, followed by resummations, yields a Dyson equation familiar from other contexts
\beq
G = G_0 + G_0 \, \Sigma \, G \, .
\eeq
Here, the time integrations run over the contour forward and then backward in time, reaching above the maximal time of operators in any considered expectation value.  The self energy $\Sigma$ can be expressed in terms of source operators $j$,
\beq
\left( i \, \frac{\partial}{\partial t_1} + \frac{{\pmb \nabla}_1^2}{2 m} \right)  \psi(1) = j(1) \, ,
\eeq
with
\beq
i \, \Sigma(1,1') = \langle \Phi | T \left\lbrace j (1) \, j^\dagger (1')   \right\rbrace    | \Phi \rangle_\text{irr} \, .
\eeq

The integral Dyson equation may be converted into an integro-differential equation through application of the differential inverse operator $G_0^{-1}$ to both sides of the equation.  For a specific order of operators in $G$ \eqref{eq:GF}, one arrives at the set of Kadanoff-Baym equations \cite{kadanoff}:
\beq
\left( i \, \frac{\partial}{\partial t_1} + \frac{{\pmb \nabla}_1^2}{2 m} \right) \, G^\lessgtr (1, 1') =  \int \text{d} 1'' \, \Sigma^+ (1, 1'') \, G^\lessgtr (1'', 1')  +
\int \text{d} 1'' \, \Sigma^\lessgtr (1, 1'') \, G^- (1'', 1') \, .
\eeq
Implicit, in the expansion and resummation leading to the Dyson equation and, thus, to the Kadanoff-Baym (KB) equations, is the presumption that the impact of any multi-particle correlations dies off with time.

In different limits, for a variety of cases of $\Sigma$, the K-B equations yield a variety of useful known results and moreover they interpolate and extrapolate beyond those limits and associated results.  Thus, when the mean field contribution dominates the self-energy, $\Sigma_\text{mf} >> \Sigma^\lessgtr$, in a highly degenerate system, the TDHF description for the system follows. When the scales of variation for the Green's functions and self-energies are significantly larger for the average than relative function arguments, then the quasiparticle approximation follows, with evolution governed by the Boltzmann equation,
\beq
-i \, G^< (1,1') \approx \int \text{d} {\pmb p} \, f({\pmb p}, 1) \, \text{e}^{i \, {\pmb p} ({\pmb x}_1 - {\pmb x}_{1'}) - i \, \epsilon_{\pmb p} (t_{1}-t_{1'}) } \, .
\eeq

Given the two limiting approaches contained in the KB equations, already employed in the practice of central nuclear reactions, it is natural to ponder a direct application of the nonequilibrium Green's functions to the reactions, covering the variety of circumstances that are possibly or clearly out of reach of those limiting approaches.  The problem with a direct solution of the KB equations, though, is the 8-dimensional nature of the functions that can easily overwhelm the current and near-future computational power.  By contrast, the TDHF equations involve 4 dimensions, or 5 if one counts the large number of orbitals, and these already tax the current power.

\section{Towards Reaction Simulations}

 General issues that must be considered when coping with nonuniform systems, described in terms of nonequilibrium Green's functions, include the space-time matrix form of the dynamics, preparation of an initial state and the abundance of matrix elements that need to be considered.  Thus, for a discretization involving sensible 50 values in any space-time direction, the number of matrix elements that would need to be considered is $50^8 = 4 \times 10^{13}$ for every function, of the order of $100 \, \text{TB}$ of data!

 In order to progress on those issues, we start \cite{rios_towards_2011} with mean-field dynamics in one dimension.  Our primary goals include assessing whether the initial state for nuclear reaction could be prepared within the same calculational scheme as the reaction simulations and whether all matrix elements of the Green's functions are equally important in calculations of the dynamics.  If many could be discarded, it might be possible to reduce the computational task for realistic calculations to an acceptable scale.

 When the self-energies $\Sigma^\lessgtr$ are suppressed, the KB equation for $G^<$ becomes
 \beq
 \left( i \, \frac{\partial}{\partial t_1} + \frac{{\pmb \nabla}_1^2}{2 m} - \Sigma_\text{mf}
\left( -iG^<(1,1) \right) \right) \, (-i)G^<(1,1') = 0 \, .
 \eeq
Here, we presume that the self energy $\Sigma_\text{mf}$ is local and depends only on local density, which is commonplace in nuclear physics.  In that situation, the kinetic and potential parts of single-particle Hamiltonian are, respectively, diagonal either in momentum or configuration space.  The equation may be solved in that situation by employing the split-operator method
\beq
\begin{split}
G^<(t_1 + \Delta t, t_{1'}) &   =  \text{e}^{-i \Delta t (K + \Sigma)} \, G^<(t_1, t_{1'}) \\
& = \left( \text{e}^{-i \Delta t \, \Sigma/2} \, \text{e}^{-i \Delta t \, K} \,  \text{e}^{-i \Delta t \, \Sigma/2} + {\mathcal O} \left( (\Delta t)^3 \right) \right)\, G^<(t_1, t_{1'}) \, ,
\end{split}
\eeq
 and by employing the Fast-Fourier-Transform (FFT) to move back-and-forth between the configuration and momentum representations.  In each representation, application of the respective portion of the evolution operator amounts to a multiplication of the Green's function by a phase factor, in itself assuring unitarity.  As a final simplification, as the mean-field self-energy depends only instantaneous values of the Green's function, the KB equation can be closed at the level of equal time-arguments of the function, i.e.\ density matrix, cf.\ Eqs.\ \eqref{eq:GF} and \eqref{eq:GF<}.

\begin{figure}
  \centerline{\includegraphics[width=.68\textwidth]{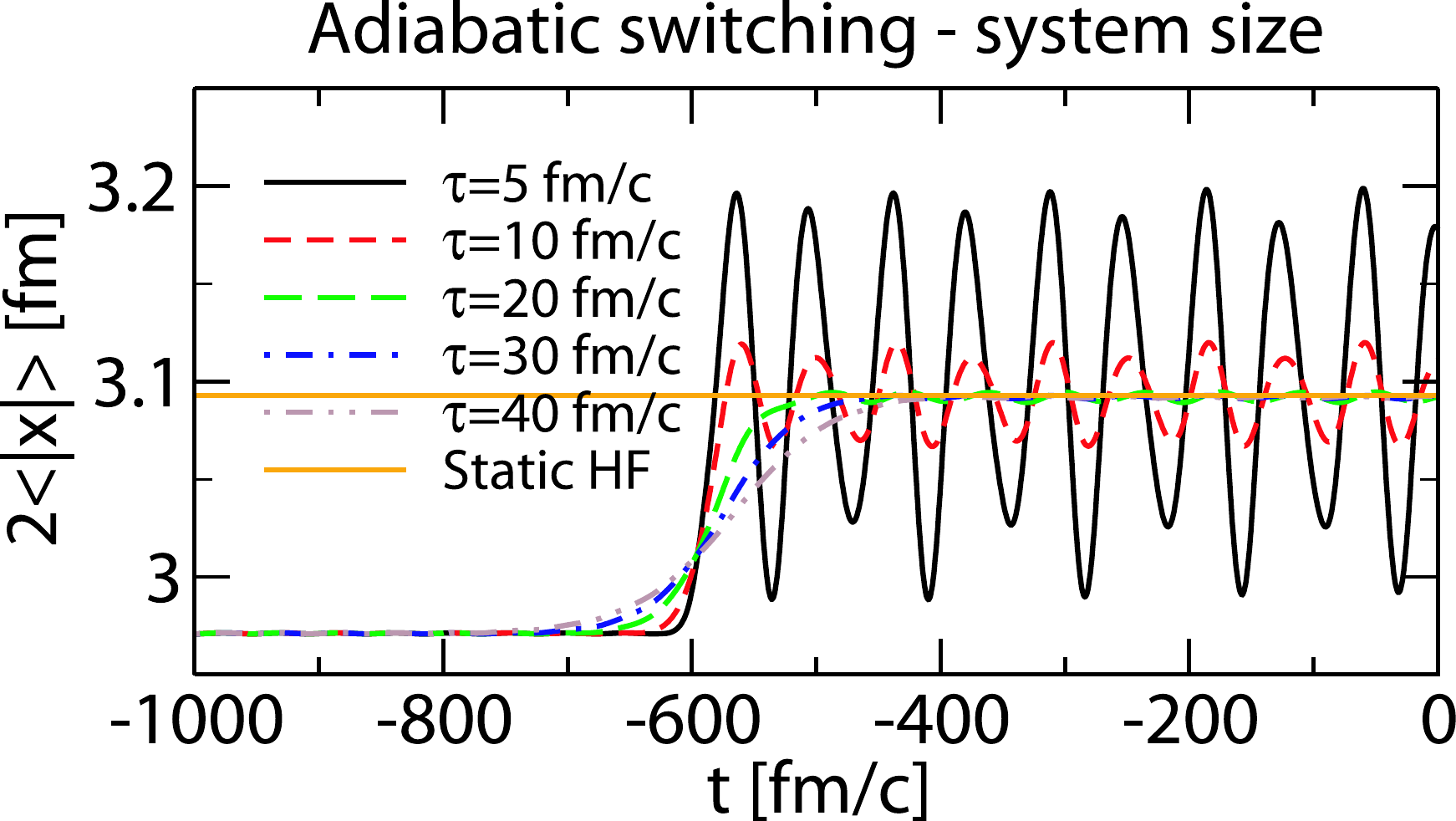}}
  \caption{Time evolution of slab width when starting from an HO configuration with $N_s=2$ filled shells at time $t=-1000\, \text{fm}/c$ and transforming the single-particle potential from the HO to the mean-field form following the Fermi-Dirac switching function with different values of the $\tau$ parameter.}
  \label{fig:SlabWidth}
\end{figure}

\begin{figure}
  \centerline{\includegraphics[width=.58\textwidth]{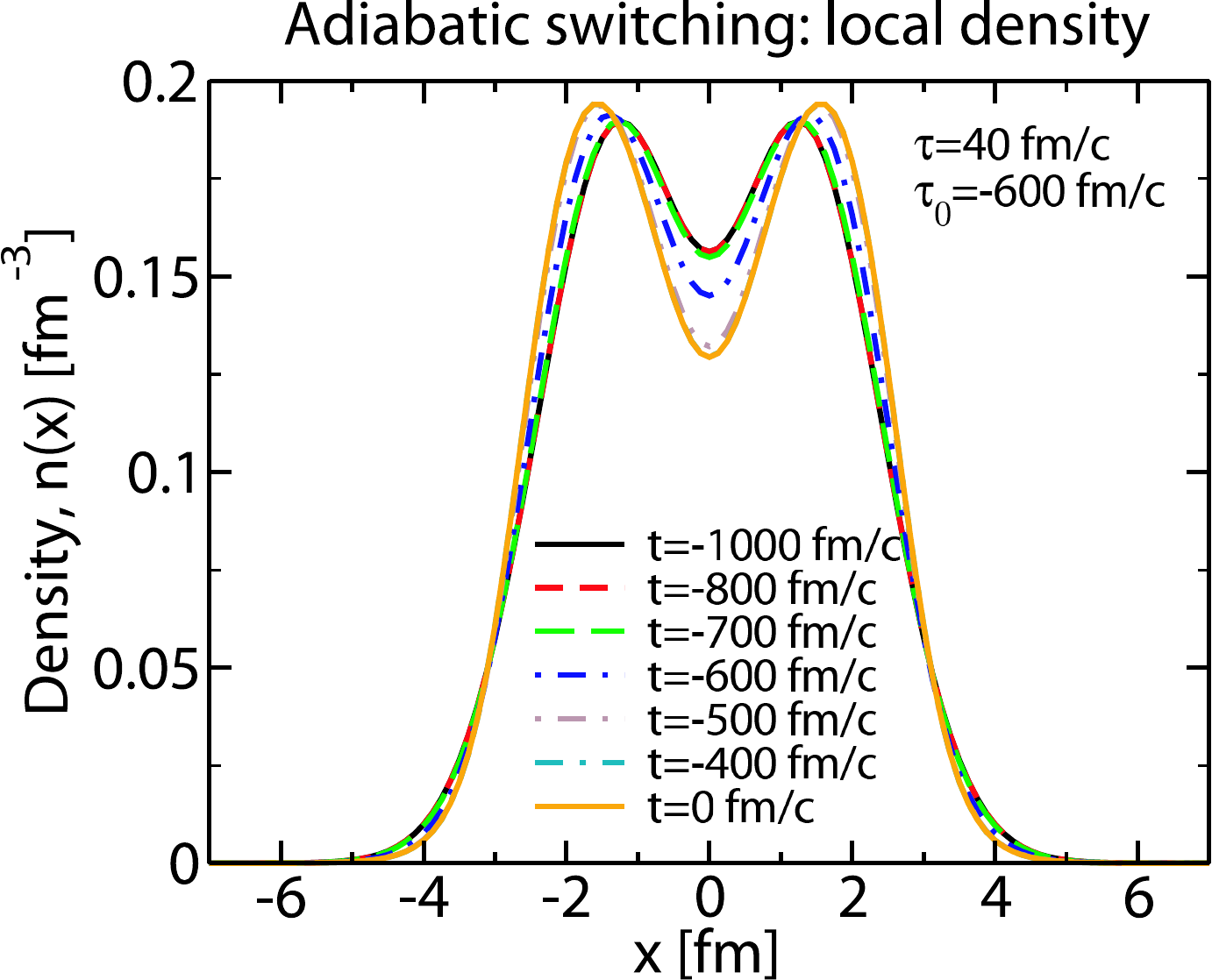}}
  \caption{Evolution of the density profile, when starting from an HO configuration with $N_s=2$ filled shells at time $t=-1000\, \text{fm}/c$ and transforming the single-particle potential to the mean-field form using $\tau = 40 \, \text{fm}/c$.}
  \label{fig:SlabProfile}
\end{figure}

 In exploring the preparation of the initial state, we test the utility of adiabatic switching of the interactions to arrive at interacting ground state~\cite{tohyama_stationary_1994,pfitzner_vibrations_1994}.  Specifically, when starting from the density matrix for a Harmonic Oscillator (HO), we evolve the Hamiltonian from that of HO to the mean-field one, using a switching profile function $f(t)$:
 \beq
 {\mathcal H}(t) = {\mathcal H}_\text{HO} \, f(t) + {\mathcal H}_\text{mf}(t) \, \left(1 - f(t) \right)  \, .
 \eeq
 The function is required to reach the limits of $f \rightarrow 1$ as $t \rightarrow -\infty$ and of $f \rightarrow 0$ as $t \rightarrow +\infty$.  Most often we use:
 \beq
 f(t) = \frac{1}{1 + \exp{\frac{t-\tau_0}{\tau}}}  \, .
 \eeq
Arriving at energy close to minimal is relatively straightforward~\cite{rios_towards_2011}.  More of a challenge is ensuring close to stationary nature of the evolved state.  Figure \ref{fig:SlabWidth} illustrates how a stationary interacting final state is gradually obtained when making the transition in the Hamiltonian more and more adiabatic.  Figure \ref{fig:SlabProfile} shows the changes in the density profile of a 1-dimensional slab when adiabatically switching over from the HO to the mean-field Hamiltonian.  We find here that it is indeed possible in practice to arrive at a satisfactory interacting ground state through adiabatic switching within the framework intended for studying reactions.  At submission of this paper, we were alerted to an exploration of the sensitivity of physics results in adiabatic switching, to switching function~\cite{watanabe_direct_1990}.

\begin{figure}
  \centerline{\includegraphics[width=.62\textwidth]{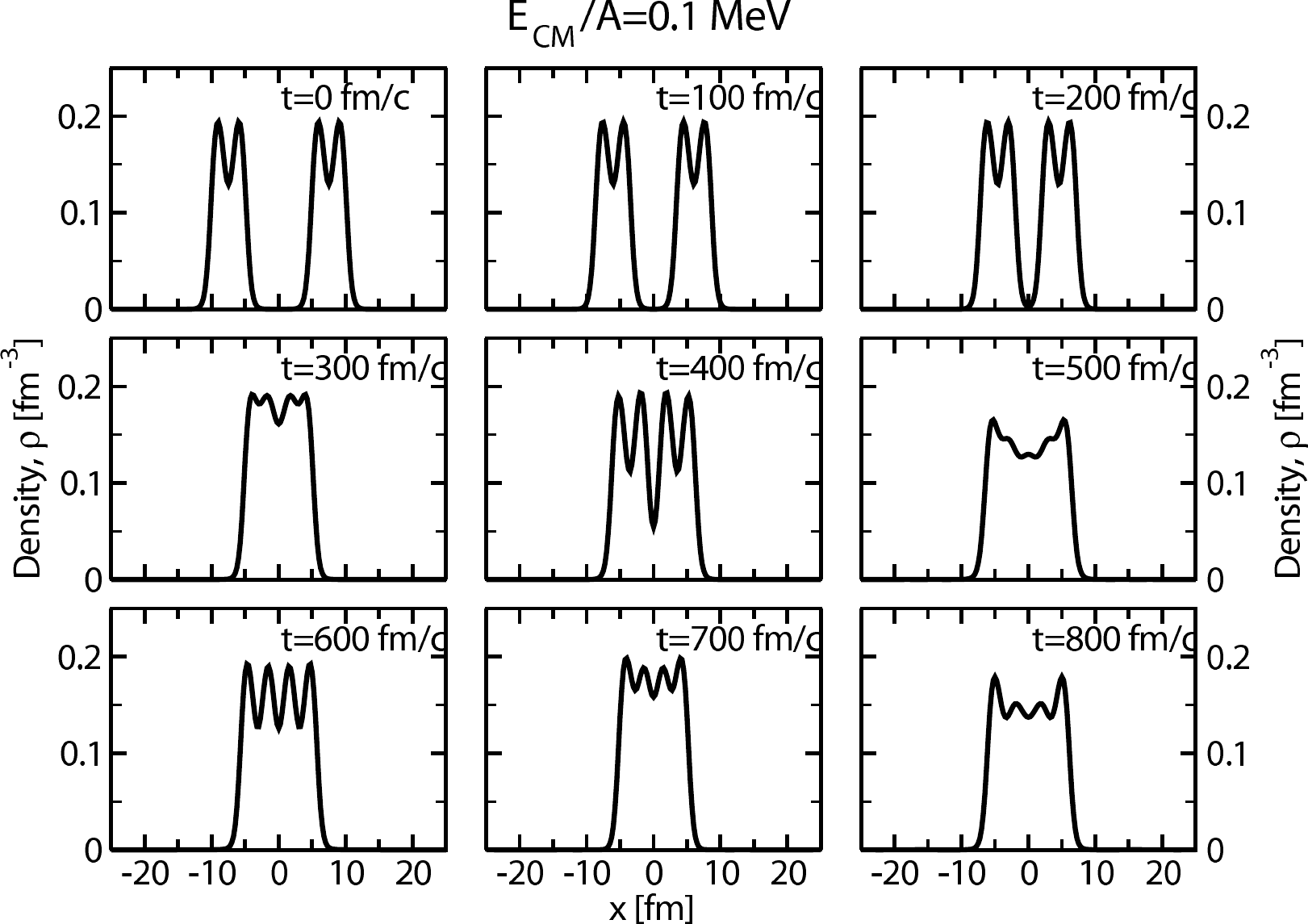}}
  \caption{Evolution of density for slabs colliding at the c.m.\ energy of 0.1~MeV/nucleon.}
  \label{fig:den01}
\end{figure}

\begin{figure}
  \centerline{\includegraphics[width=.62\textwidth]{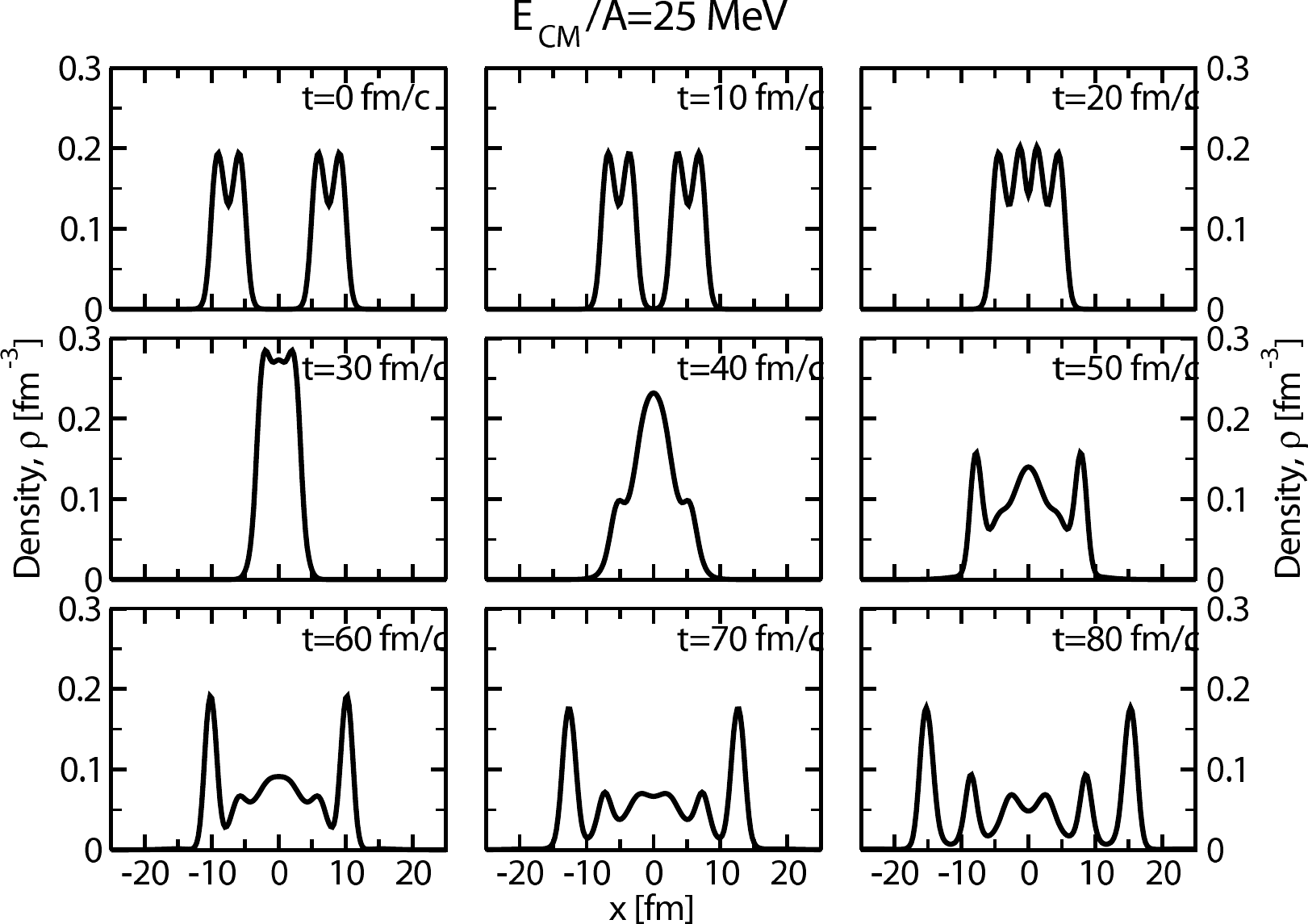}}
  \caption{Evolution of density for slabs colliding at the c.m.\ energy of 25~MeV/nucleon.}
  \label{fig:den25}
\end{figure}

We next turn to the issue of reaction dynamics and the relative importance of various elements in the density matrix and, by proxy, in the Green's functions.  To initiate a collision of slabs in one dimension, we construct the net density matrix by combining density matrices of ground-state slabs boosted to opposing momenta in the c.m.\ system:
\beq
\rho(x,x',t=0) \rightarrow
\text{e}^{ipx} \, \rho(x,x',t=0) \, \text{e}^{-ipx'} \, .
\eeq
Figure \ref{fig:den01} shows evolution of slab reaction at incident c.m.\ energy of 0.1~MeV/nucleon.  In the absence of Coulomb interactions, fusion occurs at this energy.  At energies $15 \, \text{MeV/nucleon} \gtrsim E_\text{cm} \gtrsim 0.6 \, \text{MeV/nucleon}$, the slabs pass through each other, but get excited and may break up.  At still higher energies, $E_\text{cm} \gtrsim 15 \, \text{MeV/nucleon}$, the identity of original slabs is largely lost, see Fig.~\ref{fig:den25} and the system breaks into a multitude of fragments.  In the nuclear reaction terminology, this is called multifragmentation.

\begin{figure}
  \centerline{\includegraphics[width=.65\textwidth]{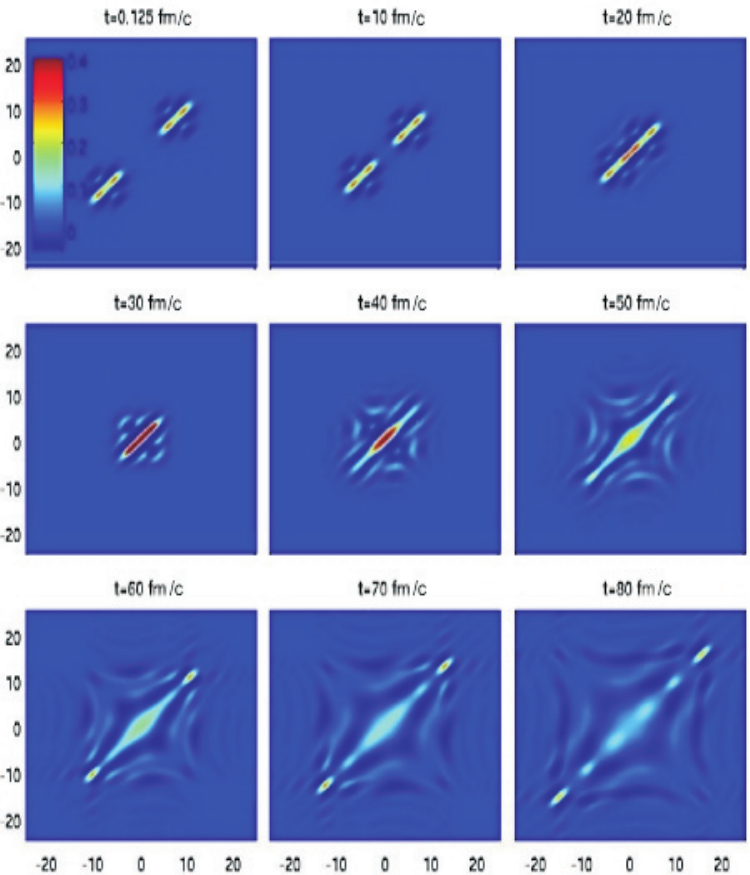}}
  \caption{Intensity plots for the real part of the density matrix $\rho(x,x',t)$ in the collision of slabs at $E_\text{cm} = 25 \, \text{MeV/nucleon}$.  The horizontal and vertical axes represent, respectively, the arguments $x$ and $x'$ of the matrix, in fm.}
  \label{fig:gxx}
\end{figure}

Figure \ref{fig:gxx} shows next the intensity plot for the real part of the density matrix for the system of slabs colliding at the c.m.\ energy of 25~MeV/nucleon.  The density matrix is always largest along the $x=x'$ diagonal and the real components tend to marginally dominate over the imaginary components.  Values of the matrix along the diagonal represent density, i.e.~Fig.~\ref{fig:den25}.  Early on in the reaction, significant values of the density matrix are limited to square-like regions of which the diagonals are colinear with the diagonal of the matrix and of which the size represents the support of nucleon wavefunctions within the individual slabs.  When the system breaks up into pieces, the region of significant values for the density matrix expands.  Besides regions adjacent to the diagonal, islands and streaks of significant values emerge, moving away, with time, from the diagonal.  These are associated with the fragmentation of the original wavefunctions, into pieces taking off with different fragments.  The regions of significant values for $\rho(x,x',t)$ represent phase correlations between amplitudes for the nucleons, from the individual original states, to move away with one or another fragment.

\begin{figure}
  \centerline{\includegraphics[width=.68\textwidth]{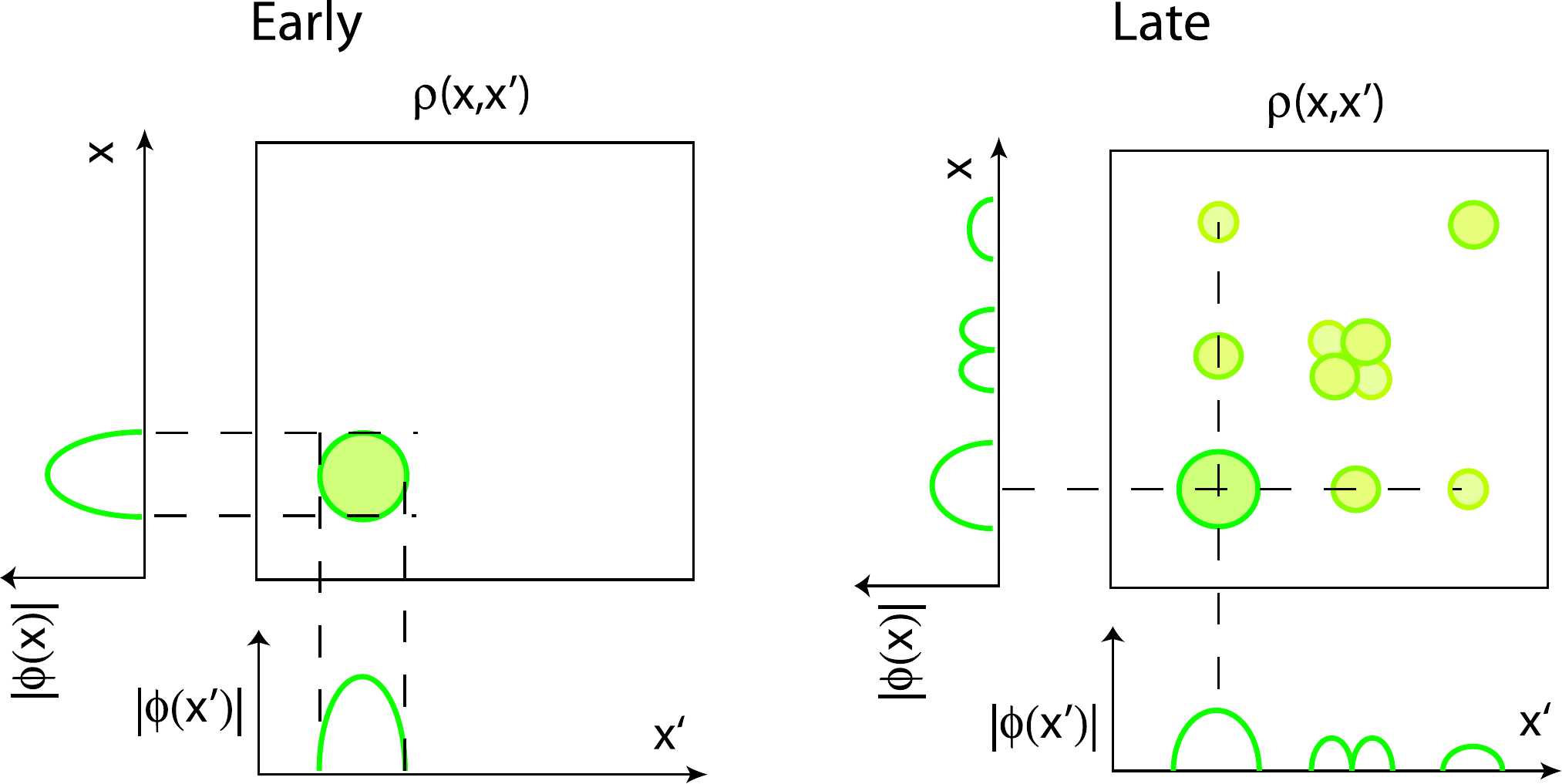}}
  \caption{Schematic illustration for the origin of significant off-diagonal elements in the density matrix $\rho(x,x',t) = \sum_\alpha n_\alpha \, \varphi_\alpha(x, t) \, \varphi_\alpha^* (x',t)$.  When nuclear fragments separate, the individual nucleon wavefunctions fragment, maintaining phase coherence between possibilities for individual nucleons moving out with one or another fragment.}
  \label{fig:rhoxx}
\end{figure}

\section{Tinkering with Evolution}

\begin{figure}
  \centerline{\includegraphics[width=.36\textwidth]{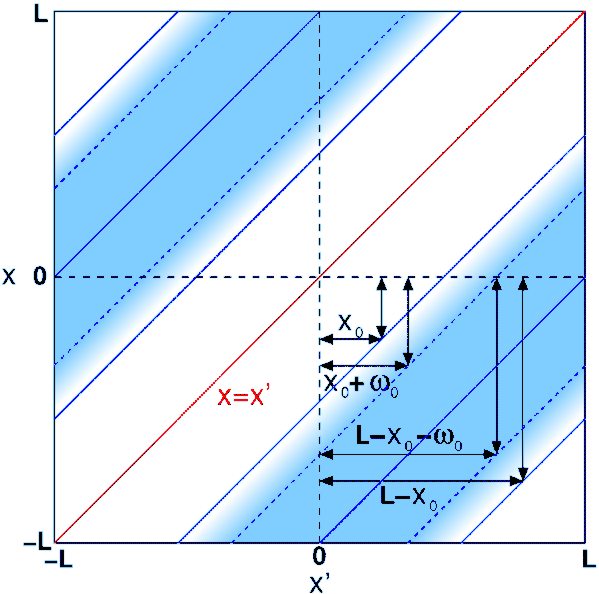}}
  \caption{Manipulation of the elements of the density matrix $\rho(x,x',t)$ in testing.  Dark regions represent elements that get suppressed.}
  \label{fig:planeg}
\end{figure}

The need to monitor far off-diagonal values of the Green's functions would have been devastating for the capability of carrying reaction simulations in 3 dimensions.  However, if the fragments from the break-up are hardly likely to ever to meet again, then the phase information in the far off-diagonal values should not matter.  To test the importance of the far off-diagonal elements, we repeat the calculations employing now a strong imaginary superoperator potential that suppresses elements away from the axis and leaves the vicinity of the axis in the density matrix intact, see Fig.~\ref{fig:planeg}.  Note that the periodic boundary conditions for the system imply a tile-like periodicity of the density matrix $\rho$ in the $x$-$x'$ plane, with the values on the diagonal repeated next on the lines passing through the the corners at $(-L,L)$ and $(L,-L)$.  With the periodicity imposed onto the superoperator, the suppression of the matrix elements occurs in valleys in the $x$-$x'$ plane, that are separated by ridges where the elements remain intact.

\begin{figure}
  \centerline{\includegraphics[width=.85\textwidth]{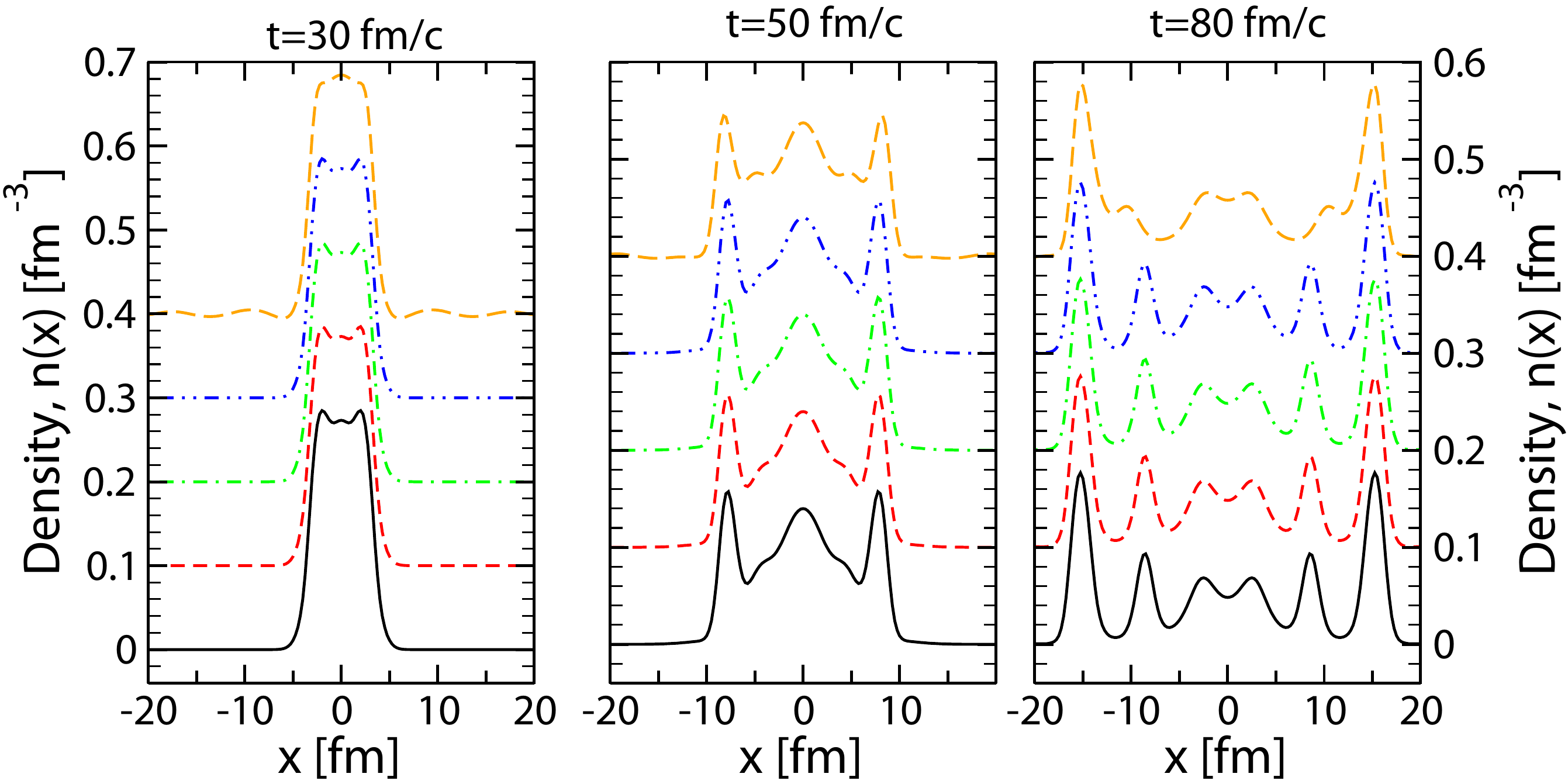}}
  \caption{Density at different stages of evolution of the system of slabs at the incident c.m.\ energy of 25~MeV/nucleon.  The bottom curves represent the density from standard evolution.  The other curves, from bottom up, represent the density obtained from evolution where the matrix elements at $|x - x'| > 20$, 15, 10, and 5~fm, respectively, are suppressed.  For clarity, the results for the density from the different evolutions are staggered by 0.1~fm$^{-3}$.}
  \label{fig:suppress}
\end{figure}

\begin{figure}
  \centerline{\includegraphics[width=.46\textwidth]{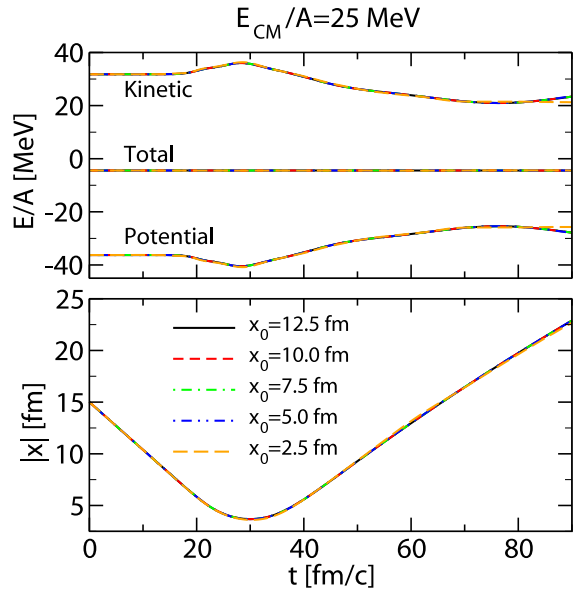}}
  \caption{Net energy components (top) and system size (bottom) for elements of density matrix suppressed at $|x - x'| > 2 x_0$, at different values of $x_0$.}
  \label{fig:energy_suppress}
\end{figure}

Figure~\ref{fig:suppress} compares results for the density at different times, obtained in the calculations of slab collisions at $E_\text{cm} = 25 \, \text{MeV/nucleon}$, for different retained regions of the matrix elements along the diagonal.  Only when the matrix elements are suppressed from $|x - x'|= 5 \, \text{fm}$ on, some changes in the density, compared to the standard evolution, begin to emerge at late times.  Even these changes are subtle as the system may still reach about the same final state, just slightly later.  Less severe trimmings of the elements leave no visible signs in the evolution of density for times relevant for reaction dynamics.  Figure \ref{fig:energy_suppress} shows further the evolution of system size and of components of net energy for different suppressions of matrix elements in the density matrix.  Only for elements suppressed at $|x - x'| > 5 \, \text{fm}$, a small change in the energy breakdown begins to emerge at late times.

Insensitivity of the dynamics to the removal of far-away matrix elements of the density matrix bodes well for applying nonequilibrium Green's functions to central nuclear reactions.  In 3 dimensions + time, such a removal would dramatically reduce the computational cost.  Importantly, it appears that one can make a compromise between the demands of low computational cost and of accuracy.  On more way of reducing the cost of calculations in 3 dimensions is in exploiting the fact that local momentum distributions are never far away from isotropy in central nuclear reactions.  With this, we hope to be able expand the functions in relative positions in terms of the cartesian spherical harmonics.  The tensorial properties of the harmonics are preserved under Fourier transformation~\cite{danielewicz_analyzing_2007} and the harmonics transform in a simple fashion under shift of the reference point~\cite{shanker_accelerated_2007}.  For weak anisotropies only a few harmonics may need to be retained.

\section{Further Advances}

\begin{figure}
			\begin{center}
					 \includegraphics[width=.10\linewidth]{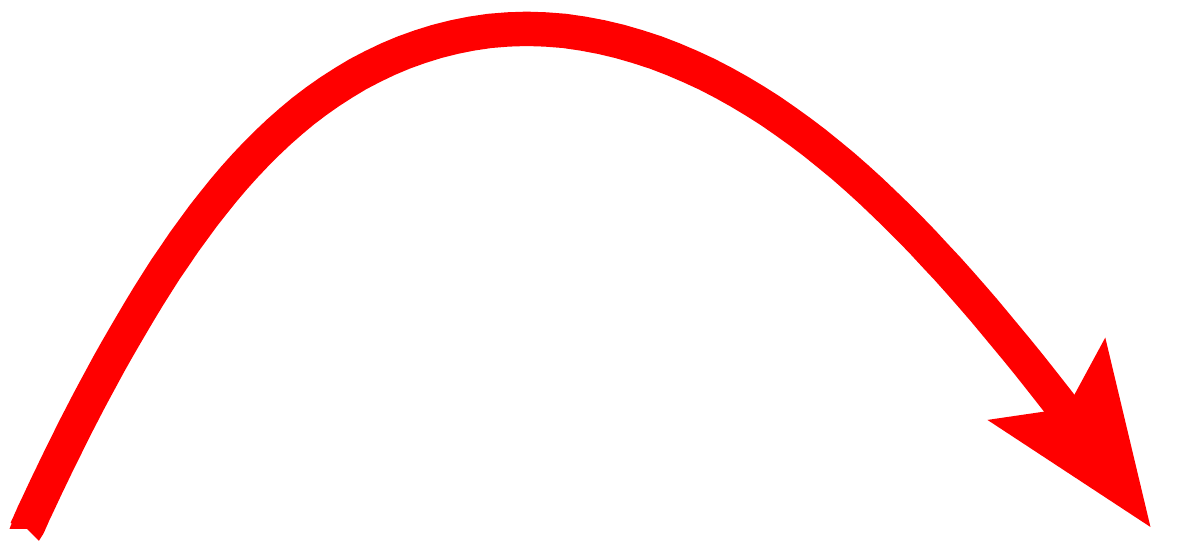}\\
						\parbox[t]{.245\linewidth}{
							 \includegraphics{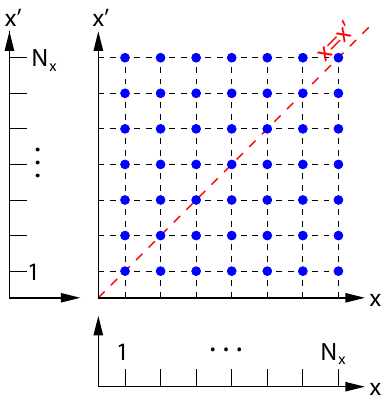}}\hspace*{1em}
						\parbox[t]{.27\linewidth}{
							\includegraphics{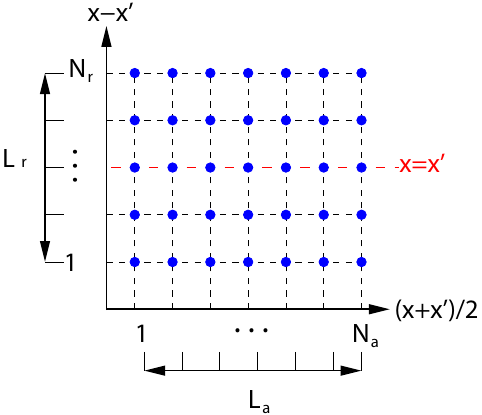}}
				\end{center}
 \caption{Change in discretization for Green's functions, when turning from the traditional to relative and average arguments.}
  \label{fig:mesh}
\end{figure}

Far away elements in the density matrix pertain to a fine momentum resolution for the Wigner function.  Elements close to the diagonal pertain to coarse characteristics in momentum.  When standard cartesian discretization employed for the Green's functions and density matrix, the momentum and position resolution scale are tied together.  However, there is no reason for coupling those.  Change of variables to relative and average arguments and discretization of the latter variables, cf.\ Fig.~\ref{fig:mesh}, allows for a decoupling of the scales and also facilitates discarding of far-away matrix elements of the functions.  The decoupling should facilitate application of the nonequilibrium Green's functions to intermediate energy central nuclear reactions, as momenta grow with energy, but scales for spatial nonuniformities stay about the same.  The rotated numerical meshes have been explored within the Ph.D.\ Thesis of B.\ Barker \cite{Barker13} and have been shown to work well.

\begin{figure}
  \centerline{\includegraphics[width=.99\textwidth]{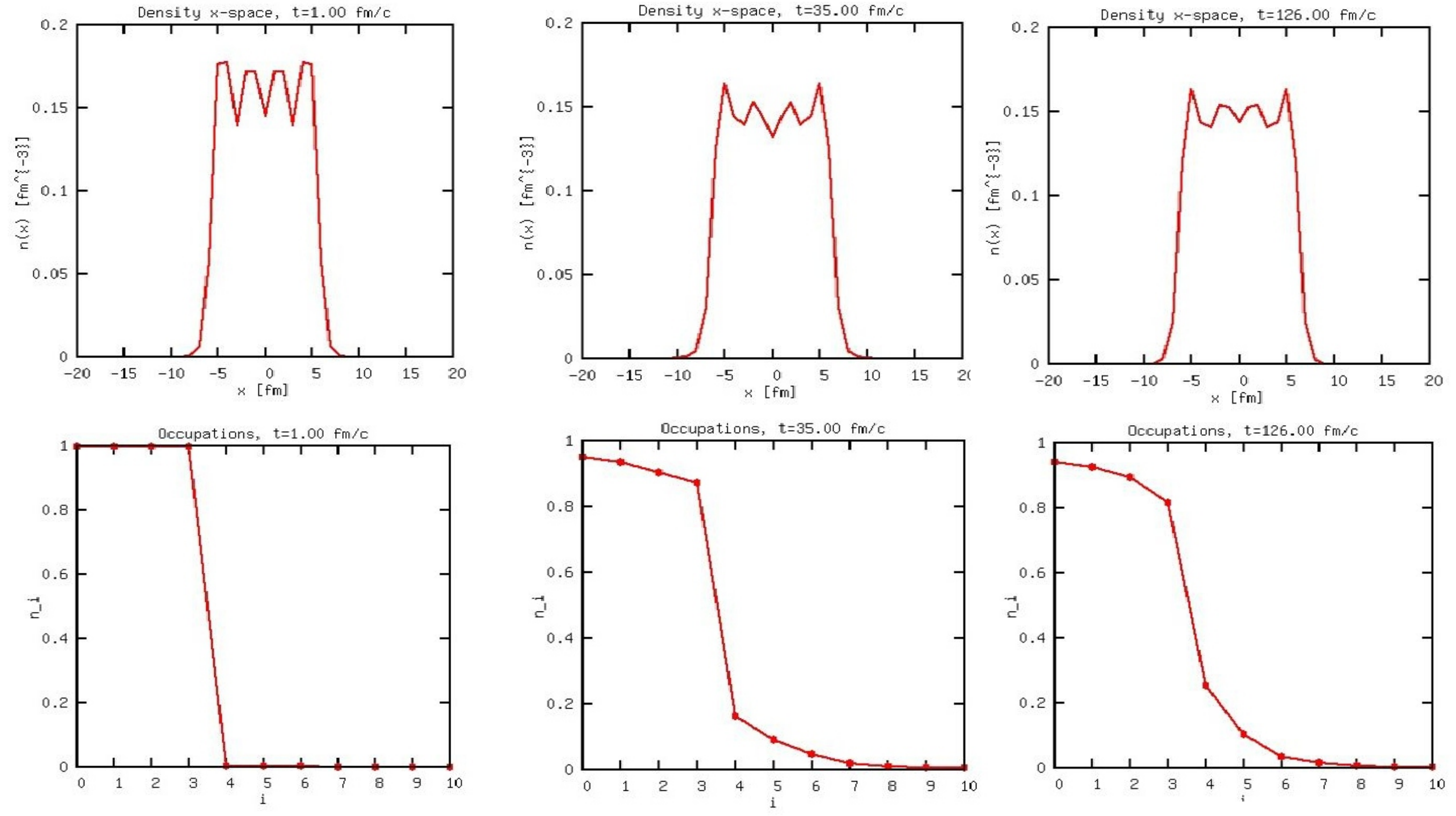}}
  \caption{Density as a function of position (top panels) and occupation (bottom), for states diagonalizing the density matrix, in a correlated evolution of an $A=16$ state starting as uncorrelated, at $t=0$, in a HO trap.}
  \label{fig:Correlated}
\end{figure}

When incorporating correlations, off-diagonal structure of the Green's functions needs to be considered \cite{danielewicz84b}, that augments the storage and calculational complexity in solving the KB equations \cite{kohler99,stan_time_2009}.  With regard to the current round of efforts, aiming at building up the infrastructure for modelling of nuclear reactions, we examined the effects of both abrupt~\cite{Kremp05} and gradual the build-up of the correlations in a homogenous and inhomogeneous system within an oscillator trap.  As an example, Fig.~\ref{fig:Correlated} shows the evolution of density in the configuration space and occupation for a slab starting in an uncorrelated state, for an abrupt build-up of the correlations.  The self-energies $\Sigma^\lessgtr$ are taken here in the Born approximation, with the characteristics of the interaction intended to reproduce nucleon-nucleon elastic scattering cross sections, in the Born approximation, at a semi-quantitative level \cite{danielewicz84b}.
Currently we concentrate on constructing correlated initial states for reaction calculations in one dimension.

\section{Conclusions}

To sum up, we have obtained encouraging results in our study of slab collisions in one dimension.  Only a~limited range of arguments in the density matrix is of practical importance for evolution.  Change of arguments in the functions facilitates dropping of far off-diagonal elements and decoupling of momentum and position resolution scales, important for application of the nonequilibrium Green's function to intermediate energy nuclear collisions.  We found that initial states for reaction calculations can be arrived at, in practice through adiabatic changes in an interaction.

\section*{Acknowledgements}

This work was partially supported by the U.S.\ National Science Foundation under Grants PHY-1068571 and PHY-1520971 by the U.K.\ Science and Technology Facilities Council under Grants ST/I005528/1 and J000051/1.

\section*{References}
%\begin{thebibliography}{9}
%\bibitem{iopartnum} IOP Publishing is to grateful Mark A Caprio, Center for Theoretical Physics, Yale University, for permission to include the {\tt iopart-num} \BibTeX package (version 2.0, December 21, 2006) with  this documentation. Updates and new releases of {\tt iopart-num} can be found on \verb"www.ctan.org" (CTAN).
%\end{thebibliography}
\bibliography{colu15}

\end{document}